\documentclass[superscriptaddress,onecolumn]{revtex4-1}
\usepackage{graphicx,caption,subcaption}
\usepackage{amsmath}
\usepackage {amssymb}
\usepackage{multirow}
\usepackage{float}
\usepackage{xcolor}
\RequirePackage[colorlinks,citecolor=blue,urlcolor=blue,linkcolor=blue]{hyperref}

\newcommand{\be}{\begin{eqnarray}}
\newcommand{\ee}{\end{eqnarray}}
\newcommand{\bea}{\begin{eqnarray}}

\newcommand{\eea}{\end{eqnarray}}

\newcommand{\dx}{ \dot{x} }

\newcommand{\Wr}{ \Omega_r }
\newcommand{\Wm}{ \Omega_m }
\newcommand{\Wl}{ \Omega_\Lambda }

\newcommand{\mnras}{\text{Mon.Not.Roy.Astron.Soc.}}
\newcommand{\km}{ \text{km}}
\newcommand{\s}{ \text{s}}
\newcommand{\Mpc}{ \text{Mpc}}

\newcommand{\LCDM}{\Lambda\text{CDM}}

\newcommand{\cecs}{Centro de Estudios Cient\'{\i}ficos (CECs), Arturo Prat 514, Valdivia, Chile}
\newcommand{\uss}{Universidad San Sebasti{\'a}n, General Lagos 1163, Valdivia, Chile}

\makeindex

\begin{document}

\centerline{\large \bf {Observational constraints on scale-dependent cosmology}}

\author{Pedro D. Alvarez}
\email{pedro.alvarez@uss.cl}
\affiliation{\cecs}
\affiliation{\uss}

\author{Benjamin Koch}
\email{benjamin.koch@tuwien.ac.at}
\affiliation{Institut f\"ur Theoretische Physik,
 Technische Universit\"at Wien,
 Wiedner Hauptstrasse 8-10,
 A-1040 Vienna, Austria  and
 Atominstitut, Stadionalle 2,
 A-1020 Vienna, Austria}
\affiliation{Instituto de F{\'i}sica, Pontificia Cat{\'o}lica Universidad de Chile, Av. Vicu{\~n}a Mackenna 4860, Santiago, Chile}

\author{Cristobal Laporte}
\email{cristobal.laportemunoz@ru.nl}
\affiliation{Institute for Mathematics, Astrophysics and Particle Physics (IMAPP), Radboud University, Heyendaalseweg 135, 6525 AJ Nijmegen,The Netherlands}

\author{\'Angel Rinc\'on}
\email{angel.rincon@ua.es}
\affiliation{Departamento de F{}sica Aplicada, Universidad de Alicante, Campus de San Vicente del Raspeig, E-03690 Alicante, Spain}


\begin{abstract}
This paper examines a cosmological model of scale-dependent gravity. 
The gravitational action is taken to be the Einstein-Hilbert term supplemented with a cosmological constant, where the couplings, $G_k$ and $\Lambda_k$, run with the energy scale $k$. 
Also, notice that, by construction, our formalism recovers general relativity when in the limit of constant Newton's coupling.
Two sub-models based on the scale-dependent cosmological model are confronted with recent observational data from: 
  i) the Hubble parameter $H(z)$, 
 ii) distance modulus $\mu(z)$, and 
iii) baryon acoustic scale evolution as functions of redshift (BAO). 
The viability of the model is discussed, obtaining the best-fit parameters and the maximum likelihood contours for these observables. Finally, a joint analysis is performed for $H(z)$+$\mu(z)$+BAO.
\end{abstract}

\maketitle

\tableofcontents

\clearpage

\section{Introduction}

Current observational evidence is compatible with a spatially flat Universe dominated by dark matter and dark energy \cite{Freedman:2003ys}. The standard cosmological model ($\Lambda$CDM hereafter) is good enough to give highly accurate predictions of the features of the Cosmic Microwave Background (CMB), primordial abundances of light elements, and large-scale structures. Although we have made significant progress, the nature of the dark sector remains an open question. $\Lambda$CDM is based on General Relativity (GR), a theory consistent with a wealth of observational data from astrophysics and cosmology. However, GR cannot be the ultimate gravity theory since some problems naturally emerge. To mention a few, the theory has the several issues: i) the existence of singularities (suggesting that GR is still incomplete), ii) the non-renormalizability, and iii) the fine-tuning problem (i.e., the issue in which fundamental constants seem to require precise values of parameters to be compatible with the observable universe).

Apart from the issues of GR mentioned above, there are a couple of tensions within the $\Lambda$CDM model itself that have been made clear by the high precision measurements of recent years. The first one to be mentioned is the weak gravitational lensing at low redshifts (see, for instance \cite{2018MNRAS.478.1244S,Mukherjee:2019wcg}) and the second one the tension between measurements of the Hubble constant $H_0$ (defined as $H(t) \equiv \dot{a}/{a}$ at the present-day $(z=0)$)
(see, for instance \cite{Ryden:2017dxw,Verde:2013wza,Bolejko:2017fos,Mortsell:2018mfj,Vagnozzi:2019ezj,Abdalla:2022yfr} and references therein). 
To be more precise, it is important to clarify that the Hubble tension arises not from a contrast between high-redshift and low-redshift data, but rather from discrepancies between predictions of $H_0$ within the $\Lambda$CDM model based on high-redshift data and measurements derived from a model-independent approach utilizing low-redshift data.
The value of the Hubble constant $H_0 [\km \ \s^{-1}\ \Mpc^{-1}] = 67.3\pm 1.1~$ (at 68\%CL, Planck TT,TE,EE+lowE+lensing+BAO) \cite{Planck:2018vyg} is now at $\sim 5\sigma$ tension with the most
recent result found by local measurements, $H_0 [\km \ \s^{-1}\ \Mpc^{-1}]= 73.1 \pm 1.1$ (at 68 CL) \cite{Riess:2021jrx}.
In this paper we focus on the following hypothesis: perhaps the assumption of time independence of the Newton constant and the vacuum energy density is no longer valid at cosmological scales and therefore an appropriate description can be given in terms of an effective action. This could mean that an improvement to the $\Lambda$CDM model could be made by including the scale dependence of the coupling constants.

Many other approaches are used to study the nature of dark energy. A phenomenological description of dark energy could use a perfect fluid characterized by a time-varying equation of state (EOS) denoted $w(a)$, where $a$ represents the scale factor. However, many other proposals based on fundamental principles do exist and can be classified into two broad categories: 
i) Dynamical dark energy models, where a new dynamical field is introduced to induce the acceleration of the universe \cite{Copeland:2006wr}, and
ii) Geometric dark energy models, where a new gravitational theory is postulated to modify Einstein's general relativity on cosmic scales \cite{Sotiriou:2008rp,DeFelice:2010aj}.
We refer to \cite{Clifton:2011jh,Nojiri:2017ncd} for comprehensive reviews.
Recently, the study of cosmological models in which the cosmological constant is assumed to be a function of a certain energy scale has attracted attention.
The reason is that a time-dependent vacuum energy makes possible to alleviate the tension in measurements of $H_0$  \cite{Poulin:2018cxd,Schoneberg:2021qvd} (see also e.g. \cite{Basilakos:2015vra,Sola:2016jky,Gomez-Valent:2018nib,SolaPeracaula:2018xsi} and references therein).
The cosmological equations in the most popular $\Lambda$-varying scenarios are generalizations of the Friedmann equations, and they offer a richer phenomenology compared to the $\Lambda$CDM model \cite{Oztas:2018jsu}. There are also some works on cosmological models with a variable Newton's constant, see e.g. \cite{Jamil:2009sr,Jamil:2010zz,Kahya:2014suo}. 

The present work focuses on a scale-dependent (SD) gravity model that, when used for cosmology, implies generalized Friedmann equations with time-dependent quantities.
\textcolor{blue}{The paper is} organized as follows:
Section \ref{SectionII} introduces the framework of the SD gravity model used in this paper. In subsection \ref{SectionIIA} we discuss the Null Energy Condition (NEC) to complement the gravitational equations, while subsection \ref{sec_subsecSD} provides the fundamental dynamical equations of cosmological SD model. Then, in section \ref{SectionIII}, we performed a likelihood analysis using $H(z)$ data (subsection \ref{SectionIIIA}), $\mu(z)$ data (subsection \ref{SectionIIIB}), BAO/CMB data (subsection \ref{SectionIIIC}), and a joint likelihood analysis (subsection \ref{SectionIIID}). To finish the paper, we summarize our main findings, and provide a short discussion in section \ref{remarks}.

\section{Framework: the scale-dependent formalism}\label{SectionII}

The effective action, $\Gamma[\varphi, k]$, where $\varphi$ represents the relevant fields in the theory and $k$ the renormalization energy scale, is a powerful tool commonly used in quantum field theory. The effective action encompasses quantum features absent in its classical counterpart that are relevant for an accurate description of the phenomena.
Moreover, the effective action has an extensive range of applications in fundamental particle physics, condensed matter physics, quantum cosmology, and quantum gravity. For a comprehensive review see \cite{Wetterich:2001kra}.
In this formalism
the coupling constants 
of the classical theory become SD quantities i.e., 
\begin{align}
    \{ A_0, B_0, C_0, \cdots, Z_0 \} \ \longrightarrow \ \{ A_k, B_k, C_k, \cdots, Z_k \},
\end{align}
where the left-hand side corresponds to the classical set and the right-hand side corresponds to the quantum-corrected SD  set of couplings.
 
Inspired by ideas of effective field theory, effective models of gravity have been proposed. The implementation of effective field theory for gravity or in the presence of gravity, is a complicated subject and, therefore, different methods and models have been proposed (see
\cite{Weinberg:1976xy,Hawking:1979ig,Wetterich:1992yh,Morris:1993qb,Bonanno:2000ep,Reuter:2001ag,Litim:2002xm,Reuter:2004nv,Bonanno:2006eu,Niedermaier:2006ns,Percacci:2007sz} and references therein). 
We will study the effective action 
\begin{equation} \label{SD_EH_action}
\Gamma[g_{\mu\nu},k] = \int 
\mathrm{d}^4x \sqrt{-g} 
\Bigg[
\frac{1}{16 \pi G_{k} }\Bigl(R-2\Lambda_{k}\Bigl) \ + \ \mathcal{L}_M
\Bigg],
\end{equation}
where $\Lambda_k$ and $G_k$ are the scale-dependent cosmological and Newton couplings, respectively.
A great advantage of working with effective quantum actions like (\ref{SD_EH_action}) is
that they are capable of incorporating some effects of quantum fluctuations at the level of equations of motion.
To obtain background solutions for this effective action
one has to derive the corresponding equation of motion.
Varying the effective action with respect to the inverse metric field, 
one obtains the corresponding Einstein field equations~\cite{Reuter:2004nv}
\begin{equation} \label{SD_EFE_vacuum}
G_{\mu\nu} =  T_{\mu \nu}^{\text{effec}}=- \Lambda_k g_{\mu \nu} - {\Delta t}_{\mu\nu},
\end{equation}
where
\begin{equation} \label{delta_t_tensor}
{\Delta t}_{\mu\nu} = G_k
\Bigl[
g_{\mu\nu}\nabla^{\alpha}\nabla_{\alpha}-\nabla_{\mu}\nabla_{\nu}
\Bigl]
G_k^{-1}.
\end{equation}
To get physical information out of those equations one has to set the renormalization scale in terms of the
physical variables of the system under consideration $k\rightarrow k(x, \dots)$. 
This process is called ``scale-setting''.
The connection between $k$ and the system's physical variables is not uniquely defined. 
Thus, by imposing a particular relation between $k$ and $\{t, r, \theta, \phi \}$, one makes a choice that affects the physical observables and their interpretation of this system. Thus, if one wishes to bypass such a disadvantage, one then should close the system following an alternative way. 
This can, for example, be achieved  by taking variations with respect to the  renormalization scale, i.e., 
\begin{align}\label{eq_SS}
\frac{\mathrm{d}\Gamma[g_{\mu \nu}, k]}{\mathrm{d}k} \Bigg |_{k=k_{\text{opt}}} = 0.
\end{align}
The solution of the above equation is called optimal scale $k=k_{opt}$, since small variations in this scale do, by construction not alter the value of the effective action $\Gamma$.
Albeit possible, the implementation of such an equation turns out to be cumbersome.
Finally, in order to close the system of equations in a consistent way, one can use the contraction $T^{\text{effec}}_{\mu \nu}\ell^{\mu} \ell^{\nu} = 0$, where $\ell^\mu$ is a null vector. This condition can be understood as an effective null energy condition, in the sense of the effective action.
In the following three sub-sections we will elaborate on the motivation, implementation, advantages and limitations of this condition.
Independent of which scale-setting condition is chosen, the symmetry of the physical system under consideration will also be applied for the scale-setting. E.g. in cosmology with a homogeneous space-time, the dynamical variable is time, which implies that $k=k(t),\; G=G(t),\; \Lambda=\Lambda(t), \dots$.

\subsection{Scale-dependent gravity with a null energy condition}\label{SectionIIA}
\label{sec_subsecSD}

Energy conditions play an essential role in many applications of general relativity, from cosmology to black-hole physics, and their importance in formulating singularity theorems. These energy conditions consist of restrictions on the stress-energy tensor. Their purpose is three-fold. Firstly, energy conditions allow us to get a sense of ``normal matter'' since they capture standard features of a different kind of matter. Secondly, all the properties of matter fields are contained in the Einstein's equations and many of their modifications (including the present work) through the stress-energy tensor. Therefore, one can analyze the resulting dynamical system without recurring to the complex behavior of the field content of the theory. Thirdly, energy conditions enable a conceptual simplification for bypassing complicated computation, as shown in the singularity theorems \cite{Penrose:1964wq,Hawking:1966sx}

This last point is one of the strongest criticisms about the range of validity of the energy conditions. Pointwise energy conditions on the stress-energy tensor are generally considered as over-simplification that are not able to capture all the features of the systems under scrutiny. An example is their application to quantum fields, where the violation of all pointwise energy conditions motivates the introduction of the quantum energy inequalities \cite{Ford:1994bj,Fewster:2012yh}, which allows a finite, possibly negative, lower bound. An intermediate step consists in the averaged energy conditions that average the components of the stress-energy tensor along suitable causal curves while preserving lower bounds to zero. It is noteworthy to remark that the validity of the energy conditions strongly depends on the contributions to the total stress-energy tensor. 

SD gravity with a pointwise NEC has been used in: 
i) cosmology \cite{Canales:2018tbn,Alvarez:2020xmk,Alvarez:2022mlf,Panotopoulos:2021heb,Bargueno:2021nuc},
ii) relativistic stars \cite{Panotopoulos:2021obe,Panotopoulos:2020zqa} and
iii) black holes \cite{Koch:2015nva,Koch:2016uso,Rincon:2018lyd,Rincon:2017goj,Contreras:2017eza}.
In this manuscript, we will continue the ideas presented in \cite{Canales:2018tbn,Alvarez:2020xmk,Alvarez:2022mlf}, with the inclusion of the pointwise NEC together with the modified Friedmann equations for the SD scenario of Einstein gravity with minimally coupled matter. The idea is the following. Given an observer on a null geodesic, the effective stress-energy tensor $\kappa_k \, T^{\text{eff}}_{\mu\nu} \equiv \kappa_k \, T_{\mu\nu} - \Delta t_{\mu\nu}$ must follow $T^{\text{eff}}_{\mu\nu} \, \ell^\mu \ell^\nu \geq 0$ \cite{Canales:2018tbn}. As discussed in \cite{Alvarez:2020xmk}, the NEC is independent of the SD cosmological constant; thus, it dictates the evolution of Newton's coupling through different energy scales. Values lower than zero are forbidden at a point \cite{Tipler:1978zz,Kontou:2020bta}, while values greater than zero falls into non-physical scenarios with negative values of the Newton constant at early or late evolution time \cite{Alvarez:2020xmk}. Therefore, the saturated energy condition 
\be\label{eq_NEC}
\Delta t_{\mu\nu} \, \ell^\mu \, \ell^\nu=0
\ee 
is well-justified. One can see the advantage of demanding the NEC for the whole $T^{\text{eff}}_{\mu\nu}$ instead of just $T_{\mu\nu}$ by looking at its geometrical counterpart: if the NEC is applied to $T^{\text{eff}}_{\mu\nu}$ and $\Delta t_{\mu \nu}$ fulfills the saturated NEC, it implies $R_{\mu\nu} \, \ell^\mu \ell^\nu \geq 0$, which means that non-rotating null geodesic congruence locally converges, ensuring that gravity is attractive for massless particles \cite{Kontou:2020bta}

To evaluate the validity of the inclusion of a NEC for closing the system of equations, one has to compare it with physical observables. In this sense, a first step was taken in \cite{Alvarez:2020xmk}, where the tension between early and late-time measurements of $H_0$ was studied for the scale-dependence correction of the classical $\Lambda$CDM model. In the present work, the idea is being put on the edge by confronting our theoretical model
with the latest Type Ia Supernova (SN Ia), Baryon Acoustic Oscillations (BAO), and CMB radiation observations. 

\subsection{The scale-dependent cosmological model}\label{sec_subsecSD}

In what follows, we will summarize the main ingredients and the differential equations which describe the cosmological background evolution of a SD Friedmann-Robertson-Lemaitre-Walker Universe
which is subject to a null energy condition in the gravitational sector~\cite{Alvarez:2020xmk}. 

Thus, rewriting the original equations in a
more convenient form, we have
\begin{align}
&\frac{1}{H_0^2}\left(H^2 - H \frac{\dot g}{g}\right)=\Wl \lambda + \frac{\Wr}{a^{4}} g + \frac{\Wm}{a^{3}} g \,,\label{SD1}\\
&\frac{1}{H_0^2}\left(2  \dot H + 3H^2-2H\frac{\dot g}{g}+2\frac{\dot g^2}{g^2} - \frac{\ddot g}{g} \right)=3 \Wl \lambda - \frac{\Wr}{a^{4}} g\,,\label{SD2}\\ 
& \frac{\ddot g}{g} -H\frac{\dot{g}}{g} - 2 \frac{\dot g^2}{g^2}=0\,,\label{NECCosm}
\end{align}
where $H=H(t)$, $g=g(t)$ and $\lambda=\lambda(t)$ are defined by
\begin{equation}\label{dimless}
 H(t) \equiv \frac{\dot{a}(t)}{a(t)}\,, 
\hspace{1cm} g(t) \equiv \frac{G(t)}{G_0}\,, 
\hspace{1cm}
\lambda(t) \equiv \frac{\Lambda(t)}{\Lambda_0}\,.
\end{equation}
As in the $\Lambda$CDM model, the density parameters $\Wm$, $\Wr$, and $\Wl$ describe the contents of the Universe at present time. The time dependence of the $\Lambda$ coupling that is implied in the scale--dependent scenario results in time-dependent dark energy density
\begin{equation}
 \Wl \lambda(t)=\frac{\Lambda(t)}{3H_0^2}\,.
\end{equation}
The value of $\Wl \lambda(t)$ is not independent degree of freedom, since it can be expressed in terms of the other functions and parameters from (\ref{SD1}) or (\ref{SD2}).

We will explore the phase space of the dynamical system given by Eqs. (\ref{SD1}-\ref{NECCosm}). For this discussion, it is convenient to use the dimensionless functions (\ref{dimless}) and
\begin{equation}
 x=\frac{a(t)}{a_0}\,. \label{xdef}
\end{equation}
The SD cosmological equations take the form
\begin{align}
 &\frac{x'(\tau)^2}{x(\tau)^2}-\frac{g'(\tau) x'(\tau)}{g(\tau) x(\tau)}= \Wm g(\tau)x(\tau)^{-3}+\Wr g(\tau)x(\tau)^{-4}+\Wl \lambda(\tau)\,,\label{eq1}\\
 &\frac{x'(\tau)^2}{x(\tau)^2}-2\frac{g'(\tau) x'(\tau)}{g(\tau) x(\tau)}+2\frac{g'(\tau)^2}{g(\tau)^2}+2\frac{x''(\tau)}{x(\tau)}-\frac{g''(\tau)}{g(\tau)}=-\Wr g(\tau) x(\tau)^{-4}+3\Wl \lambda(\tau)\,,\label{eq2}\\
 &\frac{g''(\tau)}{g(\tau)}-\frac{g'(\tau) x'(\tau)}{g(\tau) x(\tau)}-2\frac{g'(\tau)^2}{g(\tau)^2}=0\,,\label{eq3}
\end{align}
where
\begin{equation}
 t = (H_{100} h)^{-1}  \tau\,,\label{tau}
\end{equation}
and $H_{100} =100 \km \s ^{-1} \Mpc^{-1}$. When switching between the time variables $t$ and $\tau$ it is useful to have at hand the relation,
\begin{equation}
 \dx(t)/x(t) = (H_{100} h)  x'(\tau)/x(\tau)\,.\label{hdef}
\end{equation}

For later use, let us write the expression of the Hubble constant
\begin{equation}
   H_0 \equiv h \cdot H_{100}\cdot  x'(0)/x(0)\,,\label{hubbleconstant} 
\end{equation}

The evolution of this dynamical system can be determined by giving the initial conditions $x(\tau_0)$, $x'(\tau_0)$, $g(\tau_0)$ and $g'(\tau_0)$, where we denoted $\tau_0$ as the present value of the evolution coordinate $\tau$. 

The parameter $h$ will be one of the phenomenological parameters of our model. A particular cosmological model is defined by giving the values of two input parameters $\Omega_m$, $\Omega_r$ and four initial conditions  $x(\tau_0)$, $x'(\tau_0)$, $g(\tau_0)$ and $g'(\tau_0)$. For the subsequent analysis it is convenient to work with the parameters $\alpha \equiv x'(\tau_0)/x(\tau_0)$ and the time-rescaling parameter $h$.

\subsection{Why this model?}\label{sec_subsecSDsum}

Up to now we have introduced and contextualized the SD cosmological model with a NEC, culminating in 
the coupled equations (\ref{eq1})-(\ref{eq3}).
Let us now summarize the main reasons
why we believe that 
it is worthwhile to explore
this as simple toy model
for leading observable effects of quantum gravity.
\begin{itemize}
    \item Running $G$ and $\Lambda$: The variability of $g(\tau)$ and $\lambda(\tau)$ can be taken as model of the most striking features of effective quantum field theories, namely the variability of its couplings, but it also appears in a large number of models, as mentioned in the introduction.
    \textcolor{cyan}{}
    \item General covariance: Symmetries, are one of the most important guiding principle in theoretical physics. The underlying field equations of our model (\ref{SD_EFE_vacuum}) and (\ref{eq_NEC}) conserve this fundamental symmetry.
    \item Non-negative energy: The relation (\ref{eq_NEC}) avoids negative contributions to the energy content of the universe~\cite{Alvarez:2020xmk}.
    \item Positive $G$: The relation (\ref{eq_NEC}) avoids the dynamical evolution towards negative $G$ in the future.
    \item Locality: Another valuable concept of many successful physical theories is locality. Our model preserves this property.
    \item Second order: Higher-order field equations are genuinely hard to solve and tend to produce unstable solutions.
    The SD-NEC model is second order and avoids these problems.
    \item Candidate for solving the cosmological constant problem: The field equations (\ref{SD_EFE_vacuum}) and (\ref{eq_NEC}) can, without matter degrees of freedom, be solved exactly. The latter fact and other ideas were used in ~\cite{Canales:2018tbn} to propose a model to alleviate the cosmological constant problem.
    This is achieved by showing that in the SD-model the dimensionless product $\Lambda(t)\cdot G(t)$ is naturally driven to exponentially small but non-zero values during inflation.
    \item Asymptotic safety correspondence: In the pure $\Lambda$ and $G$ scenario this model provided an astonishing one-to-one correspondence to RG-flow predictions. The dimensionless product (i.e., $\Lambda \cdot G$) of this toy model can be mapped exactly to the RG-flow of asymptotic safety in the separatrix-limit~\cite{Canales:2018tbn}.
   This highly non-trivial result seems to hint towards a deeper relation between RG-time and cosmological time.
 \item Classical limit: A classical limit is typically very hard to achieve in top-down approaches to quantum gravity.
 The SD-NEC model has a well-defined classical limit, in terms of the particular solution with constant $G$ and constant $\Lambda$. 
    \item Predictivity: 
    Deviations from the above-mentioned classical limit, present the only free additional parameter of the model. This additional parameter is the initial value of $g'\neq 0$ e.g. in equation (\ref{eq3}).
    The model is thus highly predictive and due to the aforementioned second order of the equations, relatively easy to test.
    \item Phenomenological results: Our previous results, exploring the $H_0$ tension and a mapping to the statefinder approach gave very encouraging phenomenological results~\cite{Alvarez:2020xmk,Alvarez:2022mlf}.
    \item There are numerous models that implement non-constant couplings in one or the other way. Out of these, this particular model is special in the sense that it minimally implements SD couplings and the condition (\ref{eq_NEC}).
\end{itemize}

\subsection{Quantum gravity?}

Despite this rather long list of desirable properties, it is important to keep in mind that the SD-NEC model should be considered as no more than a bottom-up model with nice properties. 
It is not a full quantum gravity model
and a comparison to the first principle of quantum gravity (QG)  is limited for several reasons:
\begin{itemize}
    \item The system (\ref{SD1}-\ref{NECCosm}) does neither contain non-local~\cite{Arkani-Hamed:2002ukf,Esposito:1998hp} nor higher derivative effects~\cite{Stelle:1976gc,Fradkin:1981iu} of QG .
    \item Classical energy conditions such as (\ref{eq_NEC}) are probably broken in the quantum (gravity) regime and will eventually need to be refined~\cite{Bousso:2015mna,Grumiller:2019xna}.
    \item 
    A comparison to effective quantum gravity calculations is limited. Such quantum approaches typically only explore modifications at small distance scales. The reason for this is that
    modifications in the Infra-Red (IR hereafter) at cosmological scales are not under control due to a conceptual clash between singularity theorems~\cite{Hawking:1973uf} and the usual low curvature expansions in these approaches~\cite{Niedermaier:2006ns,Percacci:2007sz}, as explained in~\cite{Donoghue:1994dn}.
    \item 
    A related limitation is the unclear interpretation of SD and the scale-setting (such as \ref{eq_SS})  in effective QG theories, such as Asymptotic Safety~\cite{Donoghue:2019clr,Bonanno:2020bil}.
\end{itemize}
One can thus, not expect that this simple model can address all the problems of the classical theory of relativity, such as e.g. singularities. 
Keeping these limitations in mind, we are exploring the model (\ref{SD1}-\ref{NECCosm}) to a new depth by confronting its predictions with additional data sets.

\section{Observational constraints}\label{SectionIII}
Unfortunately, we have not been able to find analytical solutions to the system (\ref{eq1}) -- (\ref{eq3})  under generic initial data and density parameters, and for that reason, generic analytical predictions escape our knowledge. Our analysis also takes a practical perspective: It has not been demonstrated that the dynamics of the scale-dependent model studied in~\cite{Alvarez:2020xmk} can be mapped to certain well-known scalar-tensor theories. Therefore, it is natural to compare the model with data and seek possible insights into the allowed solution phase.

\begin{figure}[ht!]
\includegraphics[trim={0.2cm 0.2cm 0.2cm 0.2cm},clip,width=.5\linewidth]{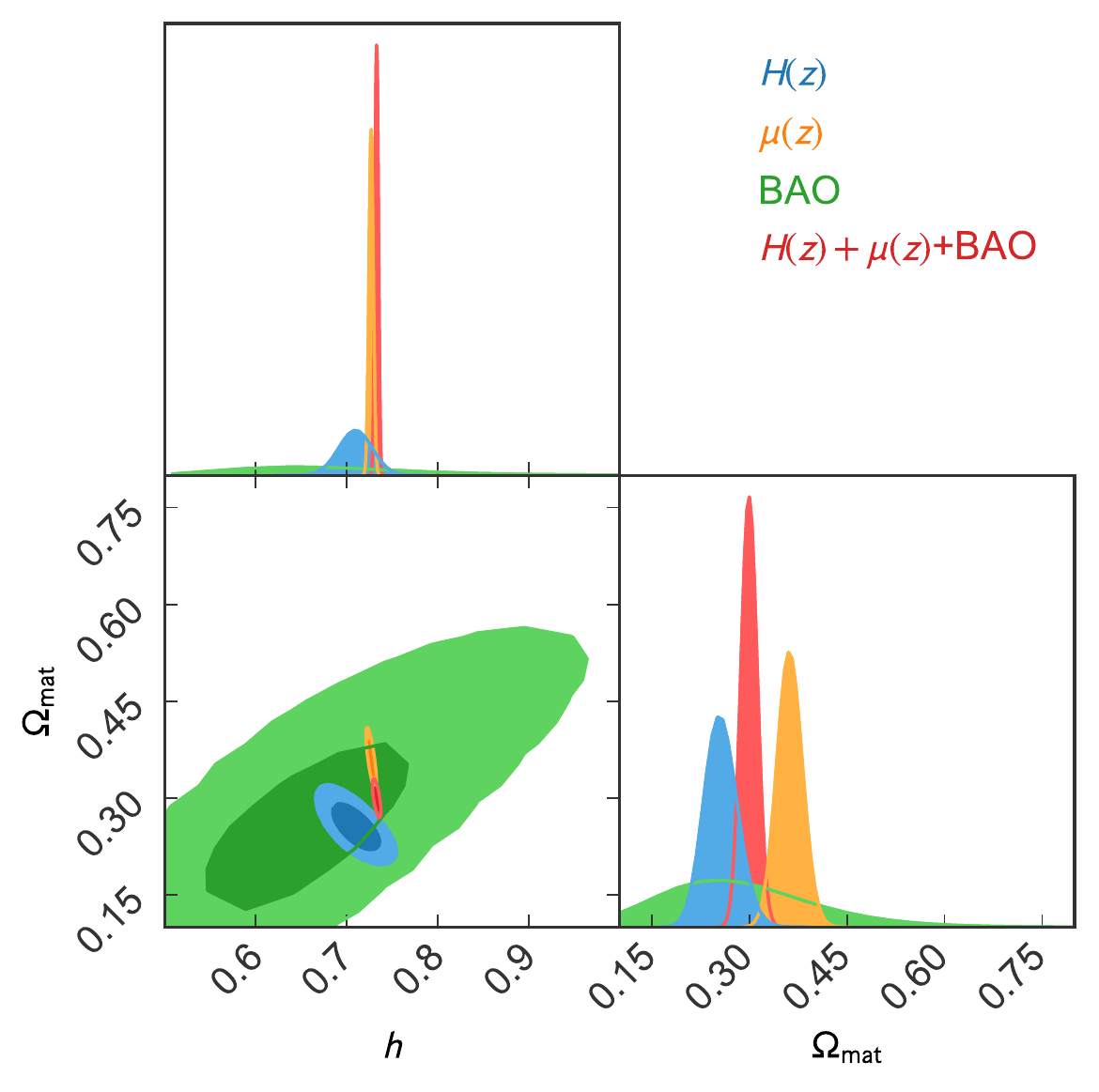}
\caption{As an illutration we included plots for $H(z)$, $\mu(z)$, BAO and the joint analysis for LCDM model.}
\label{fig:confidenceregionsLCDM}
\end{figure}

When the observables $H(z)$, $\mu(z)$ and BAO (see sections \ref{SectionIIIA}, \ref{SectionIIIB} and \ref{SectionIIIC} for a detailed explanation of these observables) are confronted to LCDM they are enough to provide a clear picture of the phenomenologically acceptable region of the free parameters, see the confidence region in the plot of fig. \ref{fig:confidenceregionsLCDM}. Then is natural to confront $H(z)$, $mu(z)$ and BAO to the scale-dependent model and respond to the question: Are these observables enough to constraint the parameters of the model?

One of the computational challenges is that, in the lack of analytical solutions, we have to use numerical solutions. Based on the results of \cite{Alvarez:2022mlf}, we integrate the equations numerically.

Based on the discussion above, we have four initial conditions $x(\tau_0)$, $x'(\tau_0)$, $g(\tau_0)$ and $g'(\tau_0)$; and two parameters $\Omega_m$, $\Omega_r$ that give us six input parameters for each cosmological model. We will trade the parameter $x(\tau_0)$ for the $h$ paramterer defined in (\ref{hdef}). For the observational data considered in the present paper (low redshift) it is fair to consider $\Omega_r = 0$.

In this paper, we will consider two main sub-models within the framework of the SD cosmology,
\begin{align}
 \theta_\text{model A} =& (x'_0, \ g'_0, \ \Wm), \quad \quad \;\,\text{with } x_0 = 1, \ g_0 = \ 1, \ \Wr=0, \, h=0.682 \,,\\
 \theta_\text{model B} =& (x'_0, \ g'_0, \ h, \ \Wm), \quad \text{with } x_0 = \ 1, \ g_0 = 1, \ \Wr=0\,.
\end{align}
The initial condition $x_0=1$ corresponds to the choice $a(\mathrm{present}) = a_0$\textcolor{red}{, see (\ref{xdef}),} while the initial condition $g_0=1$ makes sure the gravitational coupling in the present time has the observed current value\textcolor{red}{, see (\ref{dimless})}.

The value $h=0.682$ used in model A is a fiducial value based on table 8 of reference \cite{Planck:2018vyg} of cosmological parameters obtained by combining Planck TT,TE,EE+lowE with other data sets for modified gravity models that assume an effective field theory description of dark energy \cite{Cheung:2007st,Creminelli:2008wc,Gubitosi:2012hu}. The restriction of Model A helps to alleviate a high degeneration in the parameter space of Model B with respect to the observational data considered in the present paper.

We will determine the confidence regions. Our likelihood function for both sub-models is the Log-Likelihood function,
\begin{equation}
    \log P(\theta) \propto -\chi^2/2\,,
\end{equation}
where the $\chi^2$ measure is defined below for three observables separately: $H(z)$, $\mu(z)$ and BAO/CMB. 

For efficient sampling we implemented random sampling with a Markov chain Monte Carlo (MCMC) ensemble sampler using the EMCEE Python library \cite{Foreman-Mackey:2012any}. We used uniform priors and we discarded 2.5 times the largest autocorrelation time among the chains for burn-in. The number of walkers, sampling steps, burn-in steps, and thinning steps are reported in table \ref{mcmc_sampling}.
\begin{table}[h!]
\centering
\begin{tabular}{cp{1.2cm} p{1.8cm} p{1.8cm} p{1.8cm} p{1.8cm}} 
\multicolumn{2}{c}{} &  $H(z)$ & $\mu(z)$ & BAO/CMB & joint \\ \hline
\multirow{4}{*}{\begin{minipage}{12mm}
sub-model\\ A
\end{minipage}}  & $n_\mathrm{walkers}$             & 100 & 100 & 100 & 200 \\ \cline{2-6}
& $n_\mathrm{sampler}$             & 500000 & 276060 & 1000000 & 313691  \\ \cline{2-6}
                 & $n_\mathrm{burnin}$             & 107 & 106 & 491 & 112 \\ \cline{2-6}
                 & $n_\text{thin}$ & 25 & 25 & 64 & 25 \\ \hline
\multirow{4}{*}{\begin{minipage}{12mm}
sub-model\\ B
\end{minipage}} & $n_\mathrm{walkers}$             & 100 & 100 & 100 & 100 \\ \cline{2-6}
& $n_\mathrm{sampler}$               & 500000  & 687257 & 1608325 & 140000 \\ \cline{2-6}
                & $n_\mathrm{burnin}$               & 8616 & 25000 & 3004 & 11000\\ \cline{2-6}
                & $n_\text{thin}$                   & 95  & 100  & 85 & 100\\ \hline
\end{tabular}
\caption{Summary of the sampling parameters of the mcmc chains used in this paper.}
\label{mcmc_sampling}
\end{table}

In the figures, we show the $1\sigma$ and $2\sigma$ contours that in the two-dimensional case correspond to 68\%CL and 95\%CL respectively. We repeated this procedure for all the observables separately and we performed the joint analysis as well. For model A, we obtain the results of figure \ref{fig:confidenceregionsmodelA}, while for model B, the results are shown in figure \ref{fig:confidenceregionsmodelB}. The resulting mean values are summarized in table \ref{tab_results_MV}. 

\subsection{Constraints from $H(z)$ data}\label{SectionIIIA}

We can use the expressions for the redshift and the Hubble parameter,
\begin{align}
 z =& \frac{x_0}{x(\tau)}-1\,, \label{zmun}\\
 H =& h H_{100} \frac{x'(\tau)}{x(\tau)}\,.\label{Hmun}
\end{align}
in terms of the numerical solution $x(\tau)$. The subscript 0 represents present-day values and, from (\ref{xdef}), we see that it is sufficient to consider sub-models with $x_0 = 1$.

We used recent $H(z)$ data, consisting in 28+1 data points given in \cite{Rani:2014sia}. This data set is based in 28 points compiled in \cite{Farooq:2013hq} plus the value of $H_0$ estimated in \cite{Riess:2011yx}. 
To perform the likelihood analysis we define $\chi_H^2$ as
\begin{equation}
    \chi_H^2(\theta) = \sum_{i=1}^{29} \left(\frac{H(z_i;\theta)-H^\text{obs}_i}{\sigma_i}\right)^2\,.
\end{equation}
Here, $H(z_i;\theta)$ represent model predictions for a given set of parameters and
$H^\text{obs}_i$ represent the observational data points which are known with the uncertainty $\sigma_i$.

We determined best fit parameters (BF) by minimizing $\chi_H^2(\theta)$.
In the case of model A we obtained the results $x'(0) \approx 1.053$, $g'(0) \approx -1.139$ and $\Wm = 0.262$, with a value of $\chi_H^2(\theta_\text{BF}) \approx 18.779$ and a reduced chi-squared $\chi_H^2(\theta_\text{BF})/\nu \approx 0.722$, where $\nu$ is the number of degrees of freedom, $\nu(H(z))=26$ (a value $\chi^2/\nu$ around 1 shows accordance between observations and error variance). In the case of model B we obtained the results $x'(0) \approx 1.346$, $g'(0) \approx -1.398$, $h = 0.534$ and $\Wm \approx 0.436$, with a value of $\chi_H^2(\theta_\text{BF}) \approx 18.951$ and a reduced chi-squared $\chi_H^2(\theta_\text{BF})/\nu \approx 0.758$, where $\nu$ is the number of degrees of freedom, $\nu(H(z))=25$. The derived values for $H_0$ are derived using (\ref{hubbleconstant}), and we get $H_0 \approx 71.81 \km \s^{-1}\Mpc^{-1}$ for model A and $H_0 \approx 71.3 \km \s^{-1}\Mpc^{-1}$ for model B. For the sake of reference, let us report the results for $\LCDM$ as well, where $\theta^{(\LCDM)}=(H_0,\Omega_m)$. In this case we have $H_0 \approx 71.1 \km \ \s^{-1} \ \Mpc^{-1}$, $\Omega_m \approx 0.254$ with $\chi_H^2(\theta^{(\LCDM)}_\text{BF}) \approx 23.923$, $\nu = 27$ and therefore $\chi_H^2(\theta^{(\LCDM)}_\text{BF})/\nu \approx 0.886$. The $H_0$ values for Model A and Model B are somewhat higher than the result for $\LCDM$ but the difference is too small to make any remarkable claims in favour of the scale-dependent model from this point of view.

From the confidence regions (CR), we can see that the $H(z)$ data is very good at constraining model A. For model B there is degeneracy within parameters and the contours are less restrictive, see Fig. \ref{fig:confidenceregionsmodelA} and \ref{fig:confidenceregionsmodelB}. These are important leanings that will facilitate future studies of the cosmological scale-dependent model.

\subsection{Constraints from $\mu(z)$ data}\label{SectionIIIB}

The observations of type Ia supernovae (SNe Ia) directly measure the apparent magnitude $m$ of supernovae and their redshift $z$. The apparent magnitude is related to the luminosity distance $d_L$ of the supernova through
\begin{equation}
 m = M +5 \log_{10}\left( \frac{d_L(z)}{1 \Mpc} \right) + 25 \,,
\end{equation}
where $M$ is the absolute magnitude, which is considered to be constant for all SNe Ia. It is convenient to use the Hubble free luminosity distance,
\begin{equation}
 D_L(z) = \frac{H_0}{c}d_L(z)\,,
\end{equation}
and the distance modulus $\mu(z)$,
\begin{equation}
\mu \equiv m- M = 5 \log_{10}\left( \frac{D_L(z)}{H_0} \right) + 52.38 \,.
\end{equation}
The Hubble free luminosity distance $D_L(z)$ is determined using
\begin{equation}
 \frac{D_L(z)}{H_0} = (1+z) \int_0^z \frac{dz'}{H(z')}\,,
\end{equation}
where we compute $z$ and $H(z)$ using (\ref{zmun}) and (\ref{Hmun}).

For the $\mu(z)$ data, we used the Pantheon+SH0ES dataset based on nominal SN and Cepheid-host distances comprised by 1701 data points \cite{Brout:2022vxf,Riess:2021jrx}. This dataset is publicly available in the Github repository \href{https://github.com/PantheonPlusSH0ES/DataRelease}{PantheonPlusSH0ES/DataRelease}.

We define $\chi^2_\mu$ as
\begin{equation}
    \chi_\mu^2(\theta) = \sum_{i,j=1}^{1701} X_i (C^{-1})_{i,j} X_j\,, \qquad X_i=\mu(z_i;\theta)-\mu^\text{obs}_i,
\end{equation}
where $\mu(z;\theta)$ represents model predictions for a given set of parameters and $C$ is the full covariance matrix for statistical and systematic errors.

We determined best-fit parameters by minimizing $\chi^2_\mu(\theta)$. 
In the case of model A we obtained the results $x'(0) \approx 1.061$, $g'(0) \approx -0.124$ and $\Wm = 0.509$, with a value of $\chi_\mu^2(\theta_\text{BF}) \approx 1748.135$ and a reduced chi-squared $\chi_\mu^2(\theta_\text{BF})/\nu \approx 1.030$, where $\nu$ is the number of degrees of freedom, $\nu(\mu(z))=1698$. In the case of model B we obtained the results $x'(0) \approx 1.023$, $g'(0) \approx -0.120$, $h = 0.707$ and $\Wm \approx 0.473$, with a value of $\chi_\mu^2(\theta_\text{BF}) \approx 1748.135$ and a reduced chi-squared $\chi_\mu^2(\theta_\text{BF})/\nu \approx 1.030$, where $\nu$ is the number of degrees of freedom, $\nu(\mu(z))=1697$. The derived values for $H_0$ are $H_0 \approx 69.43 \km \s^{-1}\Mpc^{-1}$ for model A and $H_0 \approx 72.32 \km \s^{-1}\Mpc^{-1}$ for model B. For reference, let us report the results for $\LCDM$ as well. The best fit for $\mu(z)$ gives us $H_0 \approx 72.36 \km \ \s^{-1} \ \Mpc^{-1}$, $\Omega_m \approx 0.295$ with $\chi_\mu^2(\theta^{(\LCDM)}_\text{BF}) \approx 545.111$, $\nu = 578$ and therefore $\chi_\mu^2(\theta^{(\LCDM)}_\text{BF})/\nu \approx 0.943$.
We appreciate the fact that $\LCDM$ achieves an excellent fit with one fewer parameter.

From the confidence regions, we can see that the $\mu(z)$ data is very good at constraining $x'(0)$, but there is a large degeneracy in $g'(0)$. Model B also suffers from a large degeneracy in $h$. 
This degeneracy in $h$ is the reason why the variation of this parameter in model B did not lead to an improvement of the reduced $\chi^2$ with respect to model A, see fig. \ref{fig:confidenceregionsmodelA} and \ref{fig:confidenceregionsmodelB}. The existence of this degeneracy is also an important learning from the phenomenological point of view of scale-dependent cosmology.

\subsection{Constraints from BAO data}\label{SectionIIIC}

We can look for the parameter space of the scale-dependent model that is favored by the BAO from low-redshift data. 
The BAO scale is set by the radius of the sound horizon at the end of the drag epoch $z_d$, when photon pressure is no longer able to avoid the gravitational instability of the baryons,
\begin{equation}
 r_s= \int_{z_d}^{\infty} \frac{c_s(z)}{H(z)}dz\,,\label{rd}
\end{equation}
where the speed of sound in the photon-baryon fluid is
\begin{equation}
c_s (z) = 3^{-1/2} c {(1 + 3/4 \rho_b (z)/\rho_\gamma (z))}^{-1/2}\,, \label{cs}
\end{equation}
where $\rho_b$ is the density of barions and $\rho_\gamma$ is the density of photons.
A precise prediction of the BAO signal requires a full Boltzmann code computation, but for reasonable variations about a fiducial model, in many cases, the ratio of BAO scales is computed by the ratio of $r_s$ values computed from the integral (\ref{rd}), see for instance \cite{Aubourg:2014yra}. This often imposes a prior of the CMB measurement from the Planck satellite. We will compute (\ref{rd}) assuming dynamics from the full SD equations for $H(z)$ in (\ref{rd}). However, in the absence of the existence of a full Boltzmann study of the CMB properties within the SD cosmology, we will use the Planck values \cite{Planck:2018vyg} of the parameters in (\ref{cs}) as priors. This setup works reasonably well under the assumption that $\dot{g}_0$ is small enough, and therefore the $\Lambda$CDM predictions are close enough to the set of predictions of the SD model.

The angular diameter distance $D_A$ and the volume-averaged scale $D_V$ are related to $H(z)$ by \cite{SDSS:2005xqv}
\begin{align}
 D_A(z) =& \frac{1}{1+z} \int_0^z \frac{dz'}{H(z')}\,,\\
 D_V(z) =& \left( (1+z)^2 D_A^2(z) \frac{z}{H(z)} \right)^{1/3}.
\end{align}

We used the dataset compiled by Giostri et al for the observable $r_s(z_d)/D_V(z)$
\cite{Giostri:2012ek} that comprises: two points at $z = 0.2$ and $z = 0.35$ extracted from Percival et al \cite{SDSS:2009ocz}, a point at $z=0.106$ extracted from the 6dF Galaxy Survey \cite{Beutler:2011hx}, and three points at $z = 0.44$, $z = 0.60$ and $z = 0.73$ extracted from the BAO results by the WiggleZ team \cite{Blake:2011en}.

We define $\chi_\text{BAO}^2$ as
\begin{equation}
    \chi_\text{BAO}^2(\theta) = \sum_{i,j=1}^6 X_i (C^{-1})_{ij} X_j\,, \qquad X_i = (r_s(z_d;\theta)/D_V(z_i;\theta))-(r_s(z_d)/D_V)^\text{obs}_i
\end{equation}
where $r_s(z_d;\theta)/D_V(z_i;\theta)$ represents model predictions for a given set of parameters and $C$ is the covariance matrix~\cite{Giostri:2012ek}. 

We determined best fit parameters by minimizing $\chi^2_{BAO}(\theta)$. 
In the case of model A we obtained the results $x'(0) \approx 0.479$, $g'(0) \approx 0.408$ and $\Wm = 0.080$, with a value of $\chi_\text{BAO}^2(\theta_\text{BF}) \approx 2.159$ and a reduced chi-squared $\chi_\text{BAO}^2(\theta_\text{BF})/\nu \approx 0.720$, where $\nu$ is the number of degrees of freedom, $\nu(\mu(z))=3$. In the case of model B we obtained the results $x'(0) \approx 0.901$, $g'(0) \approx 0.768$, $h = 0.363$ and $\Wm \approx 0.282$, with a value of $\chi_\text{BAO}^2(\theta_\text{BF}) \approx 2.159$ and a reduced chi-squared $\chi_\text{BAO}^2(\theta_\text{BF})/\nu \approx 1.080$.

\begin{figure}[h]
\includegraphics[trim={0.2cm 0.2cm 0.2cm 0.2cm},clip,width=.8\linewidth]{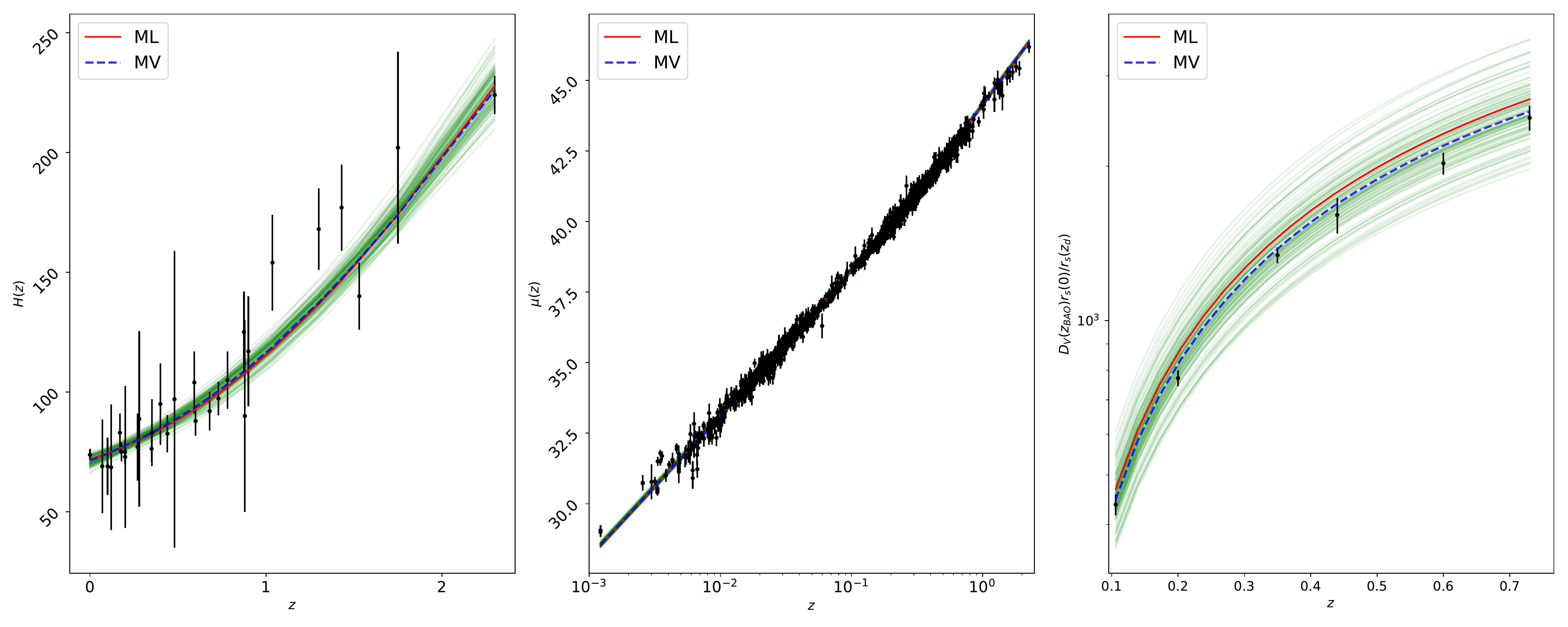}
\includegraphics[trim={0.2cm 0.2cm 0.2cm 0.2cm},clip,width=.8\linewidth]{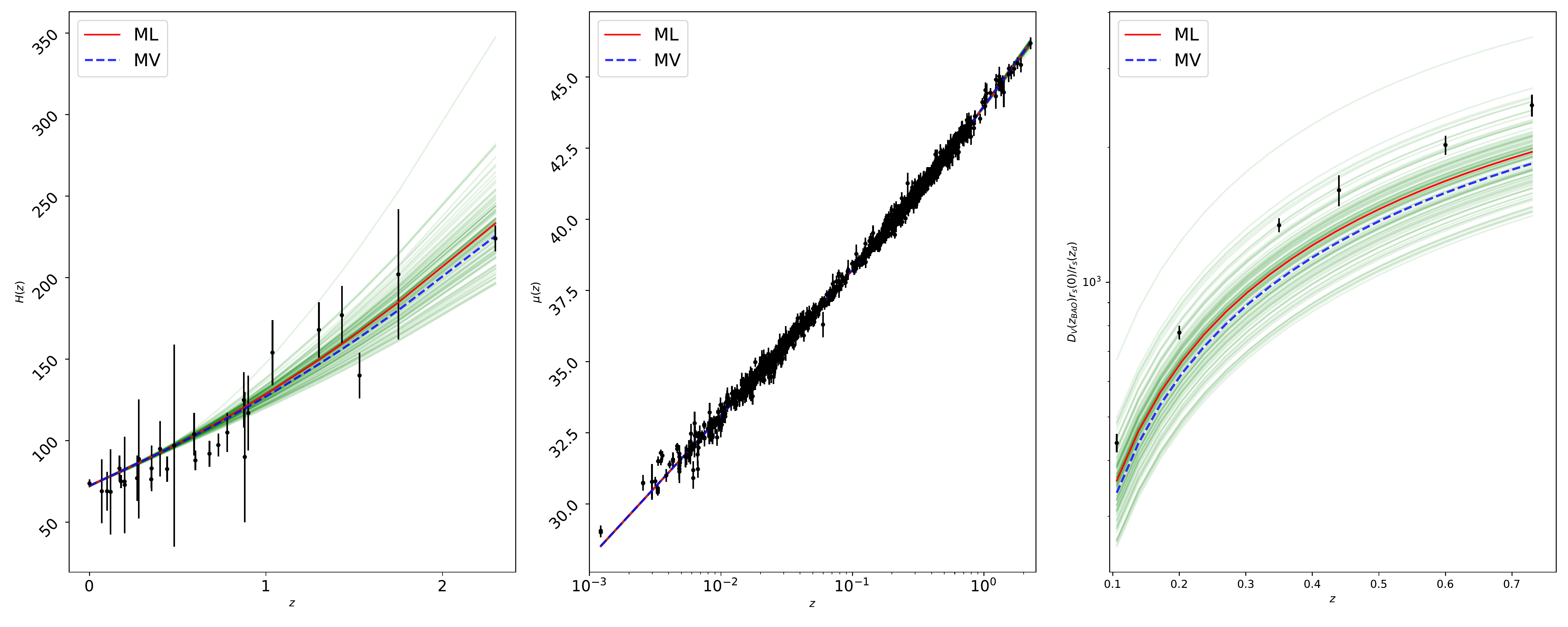}
\includegraphics[trim={0.2cm 0.2cm 0.2cm 0.2cm},clip,width=.8\linewidth]{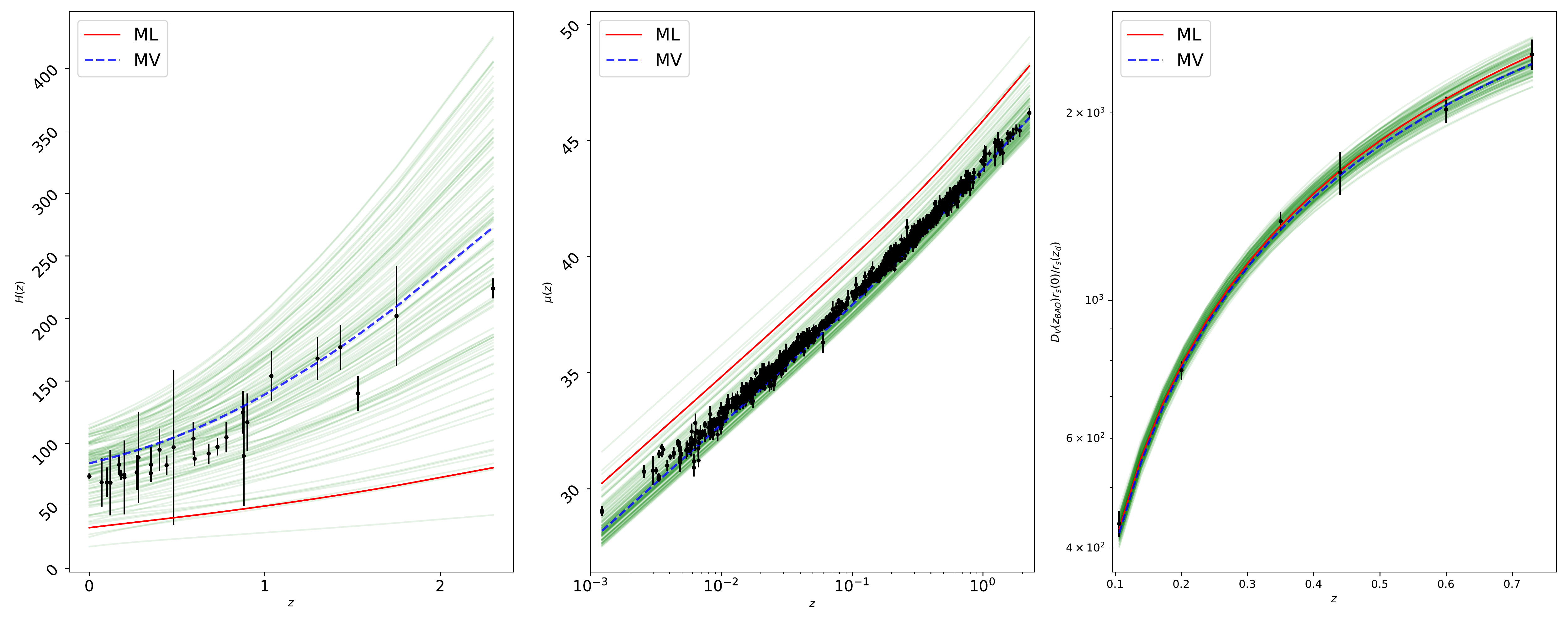}
\includegraphics[trim={0.2cm 0.2cm 0.2cm 0.2cm},clip,width=.8\linewidth]{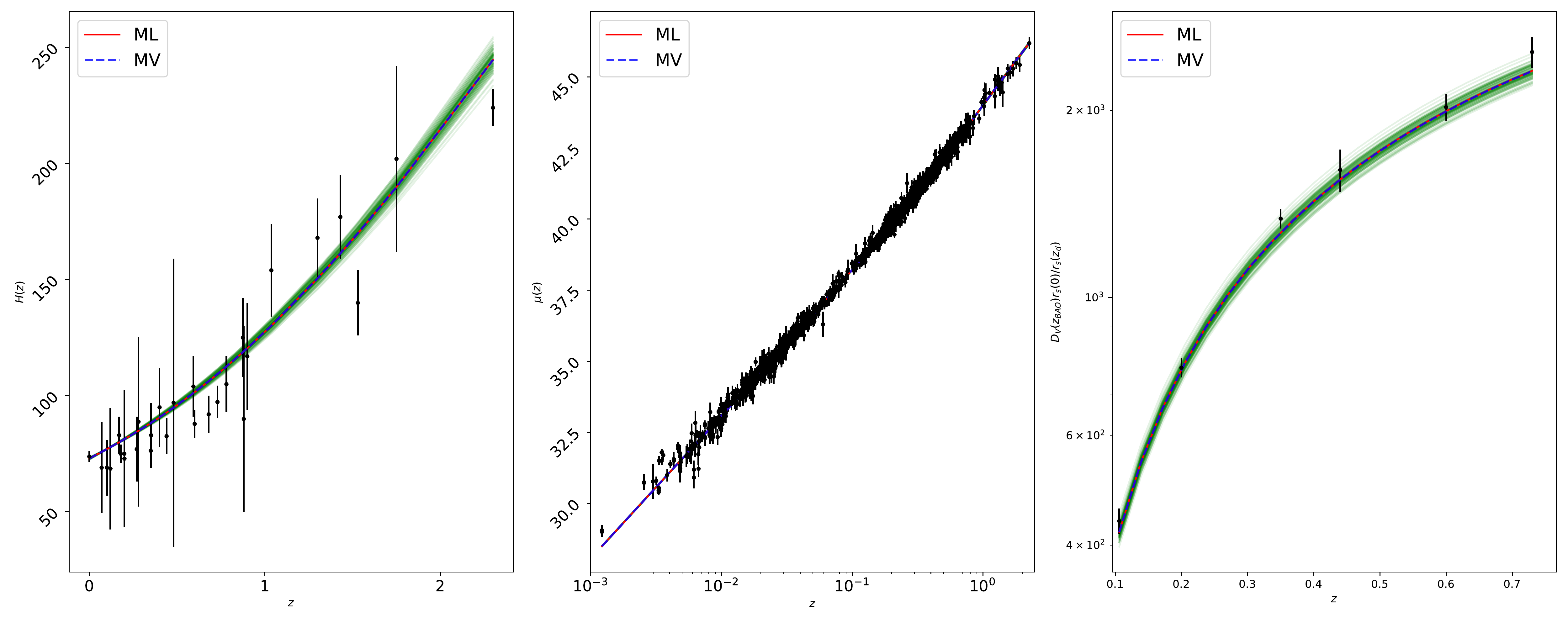}
\caption{Model A, where the parameter $h=0.682$. Red is the best-fit value (ML) and blue-dashed the mean value (MV). In green are the model predictions for a hundred randomly selected set of parameters within the posterior distribution. First row corresponds to $H(z)$ data, second row corresponds to $\mu(z)$ data, third corresponds to $BAO$ data and fourth row corresponds to joint data.}
\label{fig:BAOmodelA}
\end{figure}

\begin{figure}[h]
\includegraphics[trim={0.2cm 0.2cm 0.2cm 0.2cm},clip,width=.8\linewidth]{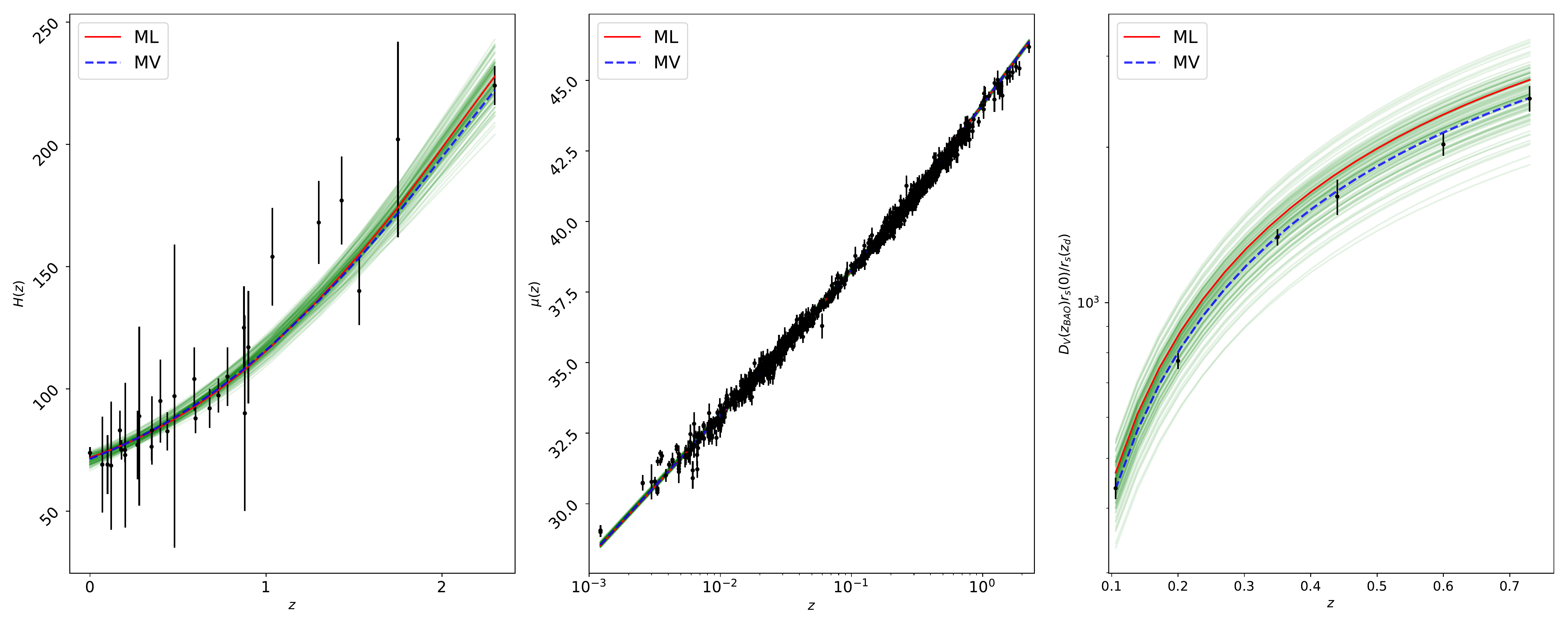}
\includegraphics[trim={0.2cm 0.2cm 0.2cm 0.2cm},clip,width=.8\linewidth]{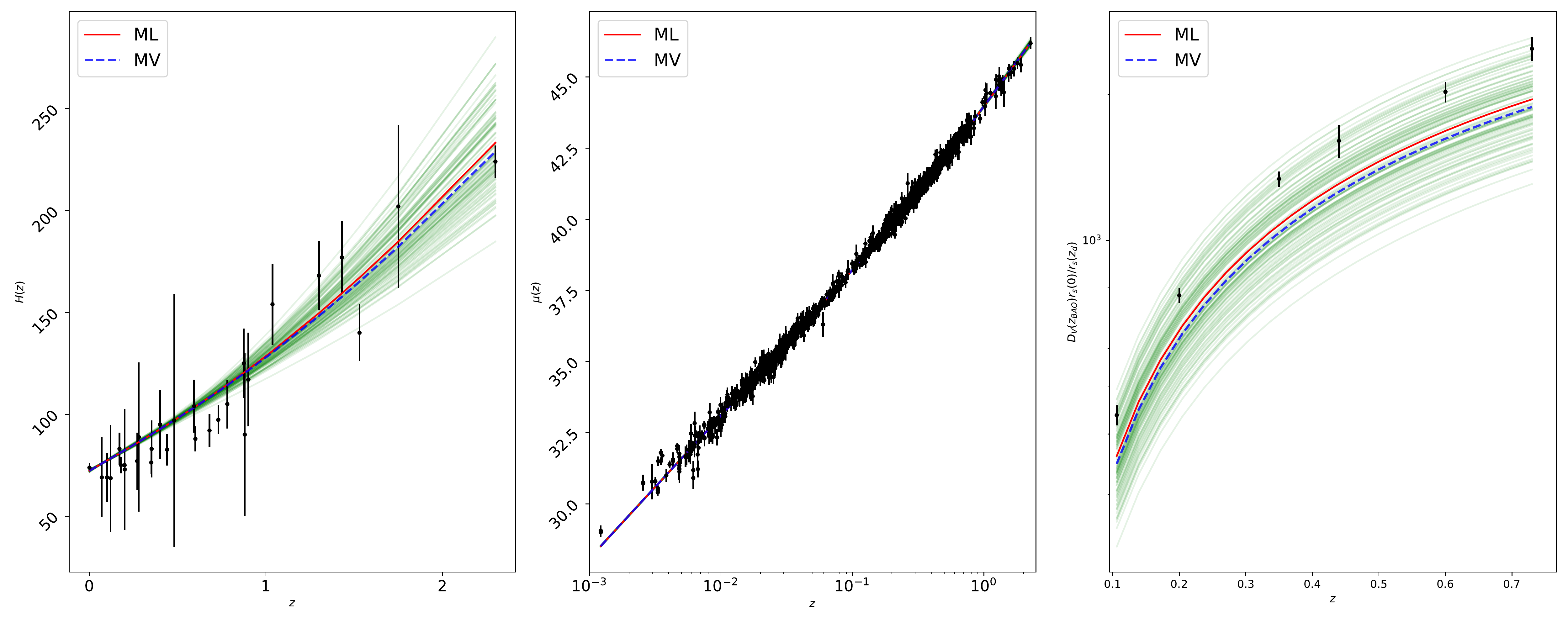}
\includegraphics[trim={0.2cm 0.2cm 0.2cm 0.2cm},clip,width=.8\linewidth]{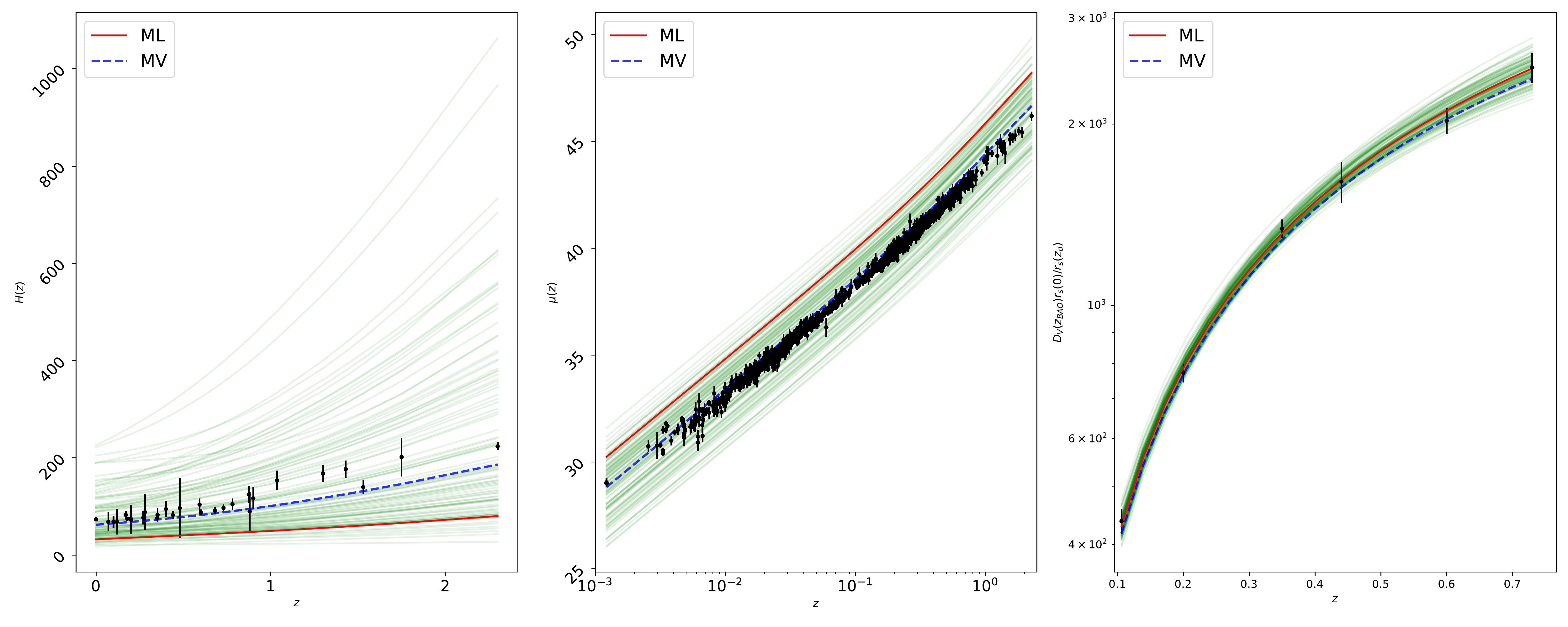}
\includegraphics[trim={0.2cm 0.2cm 0.2cm 0.2cm},clip,width=.8\linewidth]{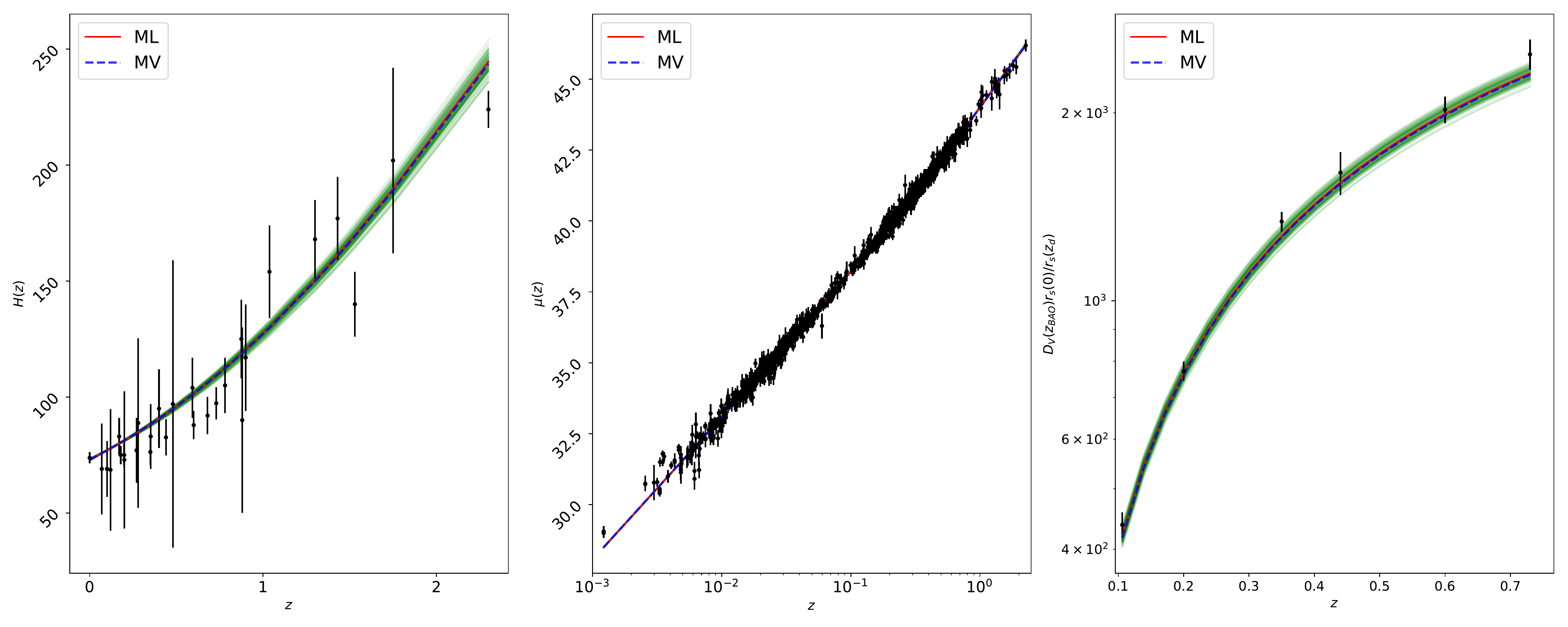}
\caption{Model B. Red is the best-fit value (ML) and blue-dashed the mean value (MV). In green are the model predictions for a hundred randomly selected set of parameters within the posterior distribution. First row corresponds to $H(z)$ data, second row corresponds to $\mu(z)$ data, third corresponds to $BAO$ data and fourth row corresponds to joint data.}
\label{fig:BAOmodelB}
\end{figure}

As it can be seen from the CR of figures \ref{fig:confidenceregionsmodelA} and \ref{fig:confidenceregionsmodelB}, there is large degeneracy for BAO within the SD model. This means that the BAO data alone is not as well suited as the other observational data sets, to constrain the parameters of the scale-dependent model. Let us note that the time dependence on cosmological scales of Newton's constant should be taken carefully because of the risk of incompatibility with large-scale structures. However, the large degeneracy that can be seen in the CR tell us that the time dependence of Newton constant implied by the scale-dependent cosmological model considered in the present paper is subtle and compatible with large-scale structure data. This can be confirmed when comparing the predicted curves for the best-fit parameters and the mean value parameters,  fig. \ref{fig:BAOmodelA} and fig. \ref{fig:BAOmodelB}. The fact that the best fit and the mean value differ substantially (see tables \ref{tab_results_MV} and \ref{tab_results_BF}) can be understood thanks to the non-Gaussianity of the contours, as it can be seen from the shape of the green contours in the CR plots.

\subsection{Joint analysis}\label{SectionIIID}

In sections \ref{SectionIIIA}, \ref{SectionIIIB}, and \ref{SectionIIIC} we constructed the CR of two sub-models based in the SD cosmology. Inspection of the 1$\sigma$ regions tell us that the observables ($H$, $\mu$, and BAO/CMB) are compatible within the framework of model A and model B. Therefore it makes sense to consider a joint analysis that could help to remove degeneracies. 

In this subsection we present the results of a joint analysis for $H(\theta)$, $\mu(z,\theta)$, and $X_i(\theta)$ of BAO/CMB.

\begin{table}[h!]
\centering
\begin{tabular}{cp{1.2cm} p{2.8cm} p{2.8cm} p{2.8cm} p{2.8cm}} 
\multicolumn{2}{c}{} &  $H(z)$ & $\mu(z)$ & BAO/CMB & joint \\ \hline
\multirow{3}{*}{\begin{minipage}{12mm}
sub-model\\ A
\end{minipage}}  & $x'(0)$             & 1.0476$^{+0.0284}_{-0.0289}$ & 1.0598$^{+0.0041}_{-0.0041}$ & 1.2339$^{+0.2693}_{-0.4432}$ & 1.0703$^{+0.0032}_{-0.0032}$  \\ \cline{2-6}
                 & $g'(0)$             & -1.0138$^{+0.3456}_{-0.2603}$ & 0.1580$^{+0.9144}_{-0.7298}$ & -1.2325$^{+0.9152}_{-0.5439}$ & -0.7955$^{+0.0729}_{-0.0704}$ \\ \cline{2-6}
                 & $\Omega_\text{mat}$ & 0.2870$^{+0.0654}_{-0.0545}$ & 0.5245$^{+0.0502}_{-0.0476}$ & 0.4124$^{+0.3061}_{-0.2412}$ & 0.4077$^{+0.0198}_{-0.0193}$ \\ \hline
                      \\ \hline
\multirow{4}{*}{\begin{minipage}{12mm}
sub-model\\ B
\end{minipage}} & $x'(0)$               & 1.4160$^{+0.3009}_{-0.4335}$  & 1.1575$^{+0.2250}_{-0.3451}$ & 1.2860$^{+0.3666}_{-0.3904}$ & 1.3425$^{+0.2401}_{-0.4021}$ \\ \cline{2-6}
                & $g'(0)$               & -1.2904$^{+0.5442}_{-0.4600}$ & 0.0511$^{+0.8932}_{-0.7477}$ & -0.8058$^{+1.5971}_{-0.8750}$ & -0.9795$^{+0.2933}_{-0.2236}$\\ \cline{2-6}
                & $h$                   & 0.5036$^{+0.2221}_{-0.0882}$  & 0.6244$^{+0.2651}_{-0.1012}$  & 0.4875$^{+0.6527}_{-0.2155}$ & 0.5436$^{+0.2324}_{-0.0823}$\\ \cline{2-6}
                & $\Omega_\text{mat}$   & 0.5290$^{+0.2953}_{-0.2845}$  & 0.6219$^{+0.2572}_{-0.3128}$  & 0.4787$^{+0.3148}_{-0.2685}$ & 0.6406$^{+0.2480}_{-0.3267}$\\ \hline
\end{tabular}
\caption{Summary of mean values obtained in this paper.}
\label{tab_results_MV}
\end{table}

\begin{table}[h!]
\centering
\begin{tabular}{cp{1.2cm} p{1.2cm} p{1.2cm} p{2.0cm} p{1.2cm}} 
\multicolumn{2}{c}{} &  $H(z)$ & $\mu(z)$ & BAO/CMB & Joint \\ \hline
\multirow{4}{*}{\begin{minipage}{12mm}
sub-\\model\\ A
\end{minipage}}  & $x'(0)$             & 1.053 & 1.061 & 0.479 & 1.070  \\ \cline{2-6}
                 & $g'(0)$             & -1.139 & -0.124 & 0.408 & -0.799 \\ \cline{2-6}
                 & $\Omega_\text{mat}$ & 0.262 & 0.509 & 0.080 & 0.407  \\ \cline{2-6}
                 &  $\chi^2/\nu$       & 0.722 & 1.030 & 0.720 & 1.039  \\ \cline{2-6}
                      \\ \hline
\multirow{5}{*}{\begin{minipage}{12mm}
sub-\\model\\ B
\end{minipage}} & $x'(0)$               & 1.346  &  1.023 & 0.901 & 1.251 \\ \cline{2-6}
                & $g'(0)$               & -1.398 & -0.120 & 0.768 & -0.934\\ \cline{2-6}
                & $h$                   & 0.534  & 0.707  & 0.363 & 0.584\\ \cline{2-6}
                & $\Omega_\text{mat}$   & 0.436  & 0.473  & 0.282 & 0.555\\ \cline{2-6}
                &  $\chi^2/\nu$         & 0.758  & 1.030  & 1.080 & 1.039\\ \cline{2-6}
\end{tabular}
\caption{Summary of the best-fit values obtained in this paper.}
\label{tab_results_BF}
\end{table}

We will assume that the observables are effectively independent by defining the combined $\chi_\text{joint}^2$ value as
\begin{equation}
 \chi^2_\text{joint} = \chi^2_H + \chi^2_\mu + \chi^2_\text{BAO}\,.
\end{equation}

In the case of model A we obtained the results $x'(0) \approx 1.070$, $g'(0) \approx -0.799$ and $\Wm = 0.407$,  with a value of $\chi_\text{joint}^2(\theta_\text{BF}) \approx 1800.367$ and a reduced chi-squared $\chi_\text{joint}^2(\theta_\text{BF})/\nu \approx 1.039$, where $\nu$ is the number of degrees of freedom, $\nu(\mu(z))=1733$.  In the case of model B we obtained the results $x'(0) \approx 1.251$, $g'(0) \approx -0.934$, $h = 0.584$ and $\Wm \approx 0.555$, with a value of $\chi_\text{joint}^2(\theta_\text{BF}) \approx 1800.367$ and a reduced chi-squared $\chi_\text{joint}^2(\theta_\text{BF})/\nu \approx 1.039$, where $\nu$ is the number of degrees of freedom, $\nu(\mu(z))=1732$. The derived values for $H_0$ are $H_0 \approx 72.97 \km \s^{-1}\Mpc^{-1}$ for model A and $H_0 \approx 73.06 \km \s^{-1}\Mpc^{-1}$ for model B. These results are summarized in table \ref{tab_results_BF}.

\begin{figure}[ht!]
\includegraphics[trim={0.2cm 0.2cm 0.2cm 0.2cm},clip,width=.7\linewidth]{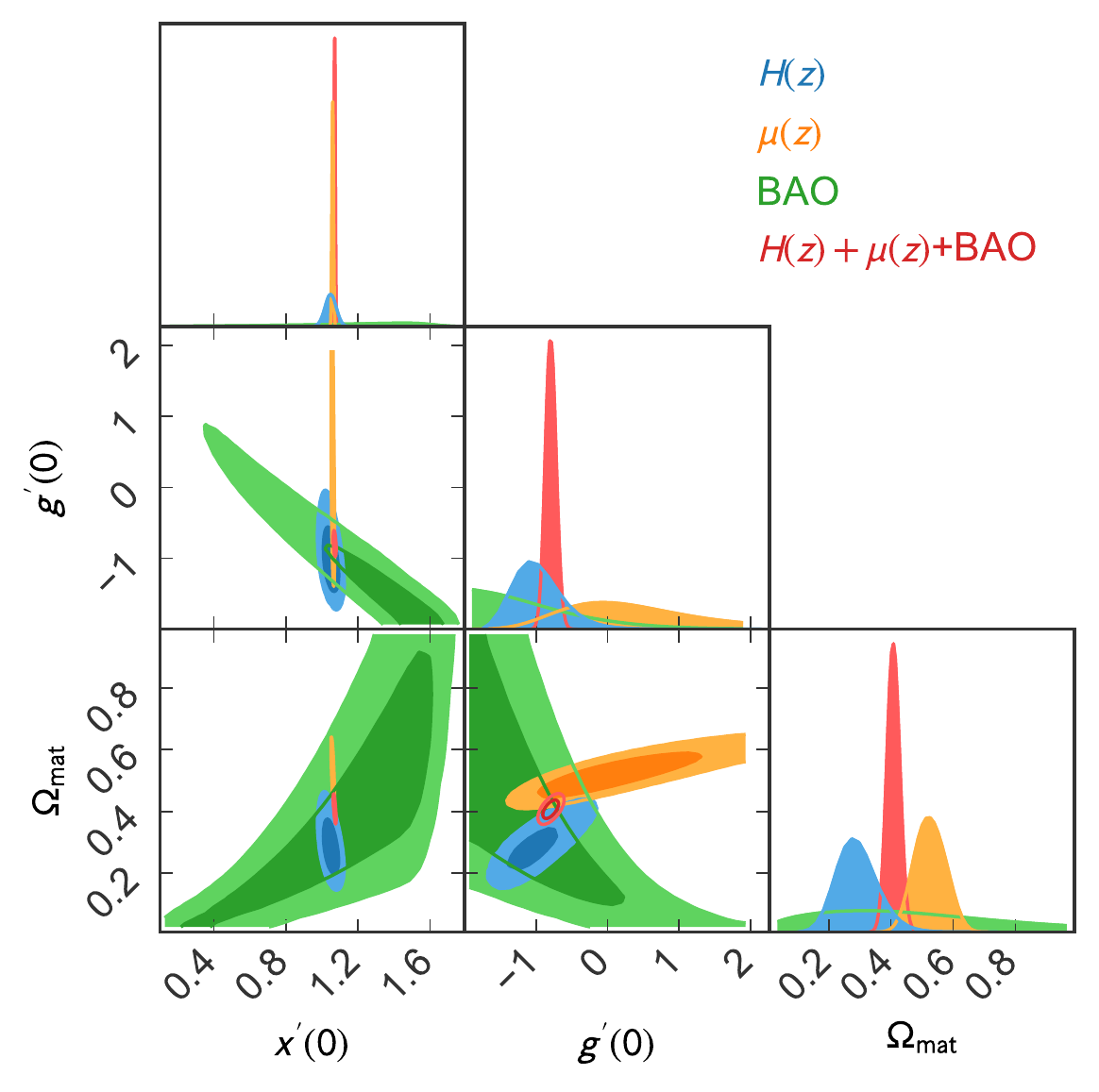}
\caption{Model A, where the parameter $h=0.682$. Corner plots showing the 1D marginalized and 2D posterior distributions. We included plots for $H(z)$, $\mu(z)$, BAO/CMB and the joint analysis. Contours for $H(z)$ show no substantial degeneracy, for $\mu(z)$ a somewhat higher degeneracy, specially in the $g'(0)$ value. The BAO/CMB contour shows large degeneracy.}
\label{fig:confidenceregionsmodelA}
\end{figure}

\begin{figure}[ht!]
\includegraphics[trim={0.2cm 0.2cm 0.2cm 0.2cm},clip,width=.7\linewidth]{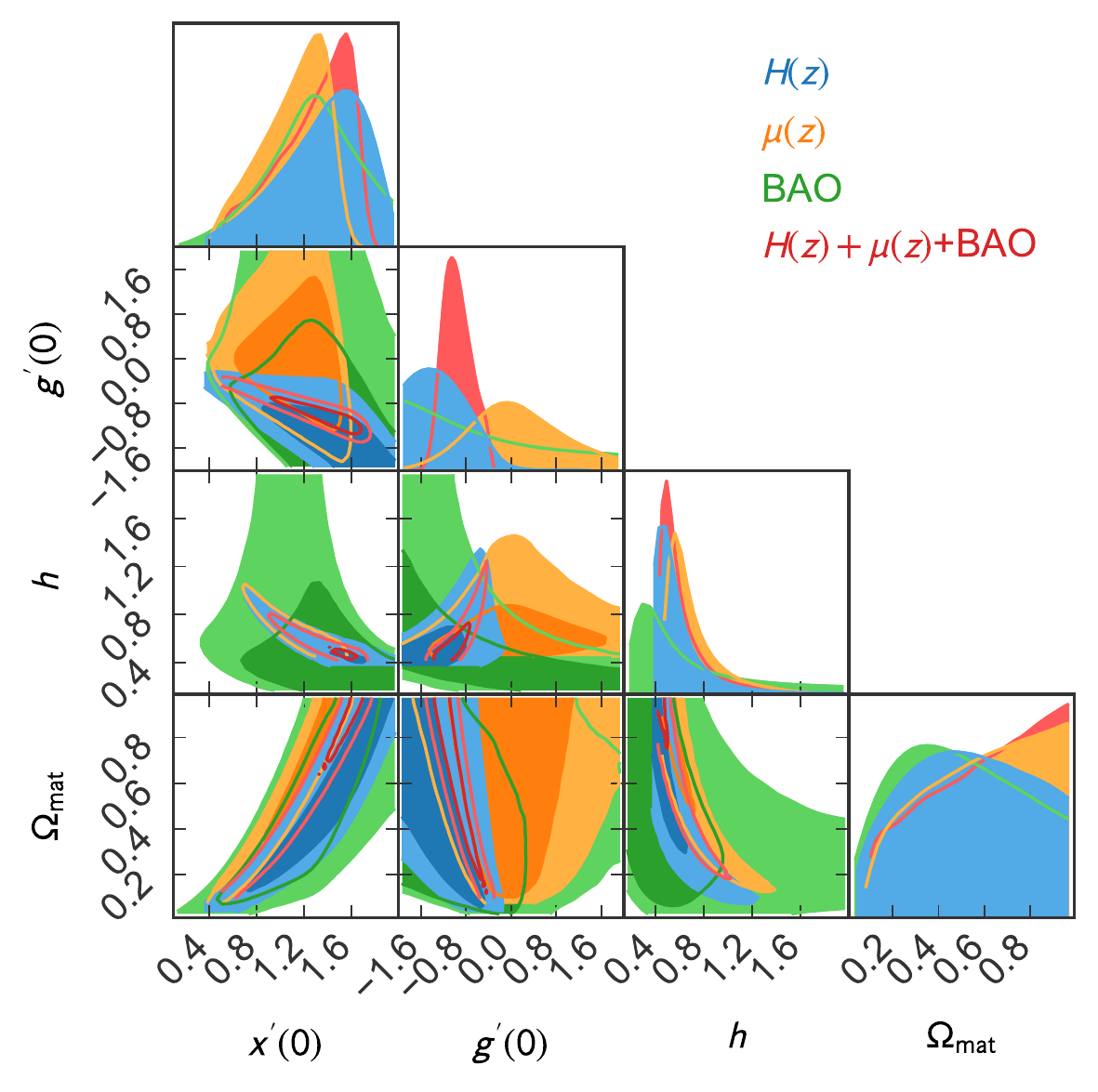}
\caption{Model B. Corner plots showing the 1D marginalized and 2D posterior distributions. We included plots for $H(z)$, $\mu(z)$, BAO/CMB and the joint analysis. The model has a large degeneracy with respecto to three observables. The joint analysis helps to reduce the degeneracy although it is not fully eliminated.}
\label{fig:confidenceregionsmodelB}
\end{figure}

The CR are depicted in the figures \ref{fig:confidenceregionsmodelA} and
\ref{fig:confidenceregionsmodelB}. The degeneracies of model A are removed, however in model B some degeneracy still remains. This can be seen as a validity check and a motivation for the study of the generation of CMB anisotropies within the scale-dependent model. These results are summarized in table \ref{tab_results_MV}.

\section{Discussion and conclusion}\label{remarks}

\begin{figure}[ht!]
\includegraphics[trim={0.2cm 0.2cm 0.2cm 0.2cm},clip,width=.4\linewidth]{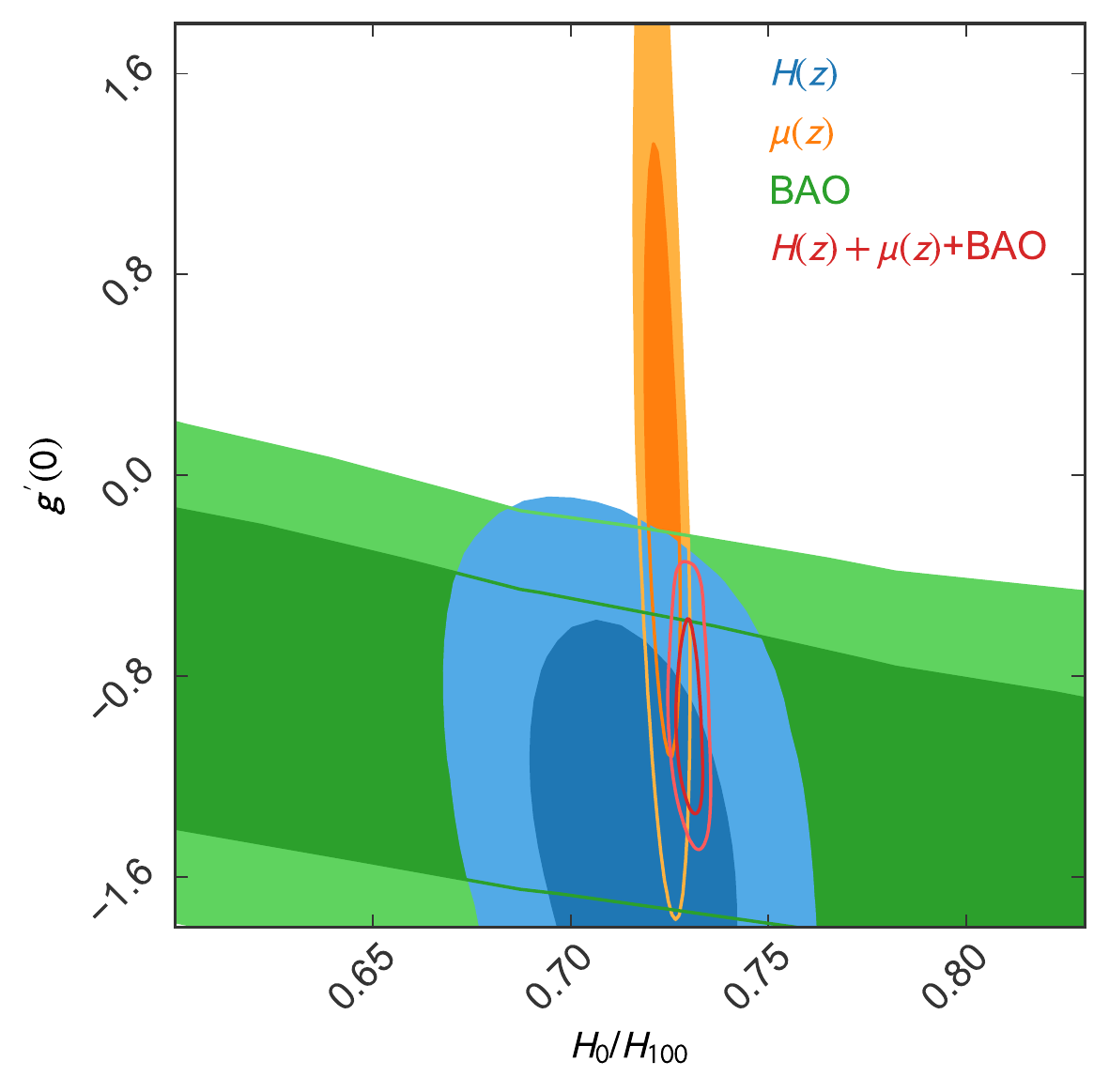}
\caption{Marginalized confidence regions of model B in terms of $g'(0)$ versus $H_0$.}
\label{fig:confidenceregiongprimeversusH0}
\end{figure}

In previous papers, we considered the SD cosmological model as a viable solution for the $H_0$ problem \cite{Alvarez:2020xmk}. In \cite{Alvarez:2022mlf} we also studied the statefinder parameters of the SD model and in particular we showed that the dynamics predicted by the SD model is not equivalent to the dynamics of Brans Dicke theory. 
In the present paper, we performed a likelihood analysis of benchmarks of the SD model,
where we allowed for three or four parameters to vary. 
We used three observables to obtain best-fit values and the maximum likelihood contours. We checked the compatibility of the observables and also performed the joint analysis.
Our present analysis is fundamental in understanding the physically relevant values of the parameters of the SD model. Let us also present marginalized results of the confidence regions of model B in terms of the parameters $g'(0)$ and $H_0$ that is the familiar expansion rate, see fig \ref{fig:confidenceregiongprimeversusH0}.

There are future refinements that can be made to our study. With respect to the sound horizon $r_d$ that is used as the standard ruler to calibrate the BAO observations \cite{Heavens:2014rja,Verde:2016ccp}, and following the ideas of \cite{Zhang:2020uan}, it is interesting to carry out a full measurement of the parameters of the SD cosmological model and the scale of the BAO from low-redshift data. It would be also interesting to implement a full Boltzmann equation solver to obtain CMB-SD predicted parameters from the CMB features only.
For the sake of simplicity, in the present model we have assumed flat spatial geometry but it would be interesting to study the implications of spatial curvature in the context of the SD model.

In summary, our analysis shows, that the presented SD model is not ruled out by the observables considered in this paper and that the best-fit values are in good agreement with the observational data sets~\cite{Farooq:2013hq,Riess:2011yx,Susuki2012,Giostri:2012ek}. It is interesting to note that the best fit and the 1$\sigma$ contour for the combined data set are compatible with a negative and non-vanishing $g'(0)$.To complement this point, let us remark that the data sets considered in the present analysis do not require to fine tune $g'(0)$ to extremely small values, as one could naively expect.

Finally, let us also remark that values of order one in $d(\ )/d\tau$ translate to values $  h \ 10^{-1} \text{Gyrs}^{-1} d(\ )/dt$ where $t$ is the physical time and those are naively in tension with bounds based on cosmological scales $|\dot{G}/G| \lesssim (10^{-4} - 10^{-2})\text{Gyrs}^{-1}$, see references therein \cite{Alvarez:2020xmk}. 
However such constraints are generically model-dependent since they typically explore a change in the gravitational coupling without taking into account e.g. the dynamic nature of such a change as it is crucial in the SD model.
Therefore our work indicates that a review of such constraints within the framework of scale-dependent gravity is timely.
Let us also clarify that we do not claim that the scale-dependent cosmology is ``better than'' any other theory, such analysis is beyond the scope of our paper and requires other statistical tools that we are not using in this paper.


\section*{Acknowlegements}

The work by C.L.\ is supported by the scholarship Becas Chile ANID-PCHA/2020-72210073. 
A. R. acknowledges financial support
from the Generalitat Valenciana through PROMETEO
PROJECT CIPROM/2022/13. 
A. R. is funded by the
Mar\'ia Zambrano contract ZAMBRANO 21-25 (Spain).
The work of P.A. is supported by Agencia Nacional de Investigaci\'{o}n y Desarrollo (ANID) through FONDECYT grant 1230112.

\bibliography{secuelaSD2.bib}

\end{document}